\documentclass[aip, rsi, reprint]{revtex4-1}

\usepackage{amsmath}

\usepackage{graphicx}
\usepackage{gensymb}

\begin{document}

\title{Single Hit Energy-resolved Laue Diffraction}
\author{Shamim Patel}
\affiliation{Department of Physics, Clarendon Laboratory, University of Oxford, Parks Road, Oxford OX1 3PU, UK}
\author{Matthew J. Suggit}
\affiliation{Department of Physics, Clarendon Laboratory, University of Oxford, Parks Road, Oxford OX1 3PU, UK}
\author{Paul G. Stubley}
\affiliation{Department of Physics, Clarendon Laboratory, University of Oxford, Parks Road, Oxford OX1 3PU, UK}
\author{James A. Hawreliak}
\altaffiliation[Now at: ]{Institute for Shock Physics, Washington State University, Pullman, Washington, 99164-2816, USA}
\affiliation{Lawrence Livermore National Laboratory, Livermore, California, 94550, USA}
\author{Orlando Ciricosta}
\affiliation{Department of Physics, Clarendon Laboratory, University of Oxford, Parks Road, Oxford OX1 3PU, UK}
\author{Andrew J. Comley}
\affiliation{Atomic Weapons Establishment, Aldermaston, Reading, RG7 4PR, UK}
\author{Gilbert W. Collins}
\affiliation{Lawrence Livermore National Laboratory, Livermore, California, 94550, USA}
\author{Jon H. Eggert}
\affiliation{Lawrence Livermore National Laboratory, Livermore, California, 94550, USA}
\author{John M. Foster}
\affiliation{Atomic Weapons Establishment, Aldermaston, Reading, RG7 4PR, UK}
\author{Justin S. Wark}
\affiliation{Department of Physics, Clarendon Laboratory, University of Oxford, Parks Road, Oxford OX1 3PU, UK}
\author{Andrew Higginbotham}
\altaffiliation[Now at: ]{York Plasma Institute, University of York, Heslington, YO10 5DD, UK}
\affiliation{Department of Physics, Clarendon Laboratory, University of Oxford, Parks Road, Oxford OX1 3PU, UK}

\date{\today}

\begin{abstract}
\emph{In-situ} white light Laue diffraction has been successfully used to interrogate the structure of single crystal materials undergoing rapid (nanosecond) dynamic compression up to megabar pressures. However, information on strain state accessible via this technique is limited, reducing its applicability for a range of applications. We present an extension to the existing Laue diffraction platform in which we record the photon energy of a subset of diffraction peaks. This allows for a measurement of the longitudinal and transverse strains \emph{in-situ} during compression. Consequently, we demonstrate measurement of volumetric compression of the unit cell, in addition to the limited aspect ratio information accessible in conventional white light Laue. We present preliminary results for silicon, where only an elastic strain is observed. VISAR measurements show the presence of a two wave structure and measurements show that material downstream of the second wave does not contribute to the observed diffraction peaks, supporting the idea that this material may be highly disordered, or has undergone large scale rotation.
\end{abstract}

\maketitle

\section{Introduction}

Laser compression of matter has enabled the investigation of a large range of high pressure states in condensed matter materials. Such conditions are relevant across a variety of fields including planetary structure\cite{Swift2012, Schneider2011}, evolution and impact\cite{Kraus2012}. Indeed, recent advancements in quasi-isentropic compression experiments have enabled the generation of conditions consistent with those found in the cores of Jovian planets\cite{Bradley2009, Smith2014, Ping2013}, well beyond the reach of conventional static compression techniques\cite{Dubrovinsky2012}. In addition, a full understanding of the dynamical response of materials subject to laser compression is required to fully model the initial stages of capsule implosions in inertial confinement fusion \cite{Edwards2013}.

In parallel, we are also seeing a rapid evolution in the ability to diagnose bulk, structural and microstructural properties of this laser compressed matter. Such techniques include measurements of wave profiles under compression\cite{Eggert2008,Bradley2004}, analysis of recovered samples \cite{Lu2013} and \emph{in-situ} x-ray absorption\cite{Ping2013}. Of particular interest is \emph{in-situ} x-ray diffraction, which has been used to investigate the structure of materials in laser compression experiments. Techniques exist to probe both single and polycrystalline samples, with phase transitions and plasticity having been observed in a growing number of materials\cite{Kalantar1999,Milathianaki2013,Murphy2010a,Rygg2012,Coppari2012}. 

One such method is \emph{in-situ} Laue diffraction. In this technique a laser generated broadband x-ray source is collimated, and the x-rays allowed to diffract from a compressed single crystal. The resulting Laue spots are recorded on image plate based detectors, with their positions related to the orientation of the diffracting lattice planes\cite{Suggit2010}. This ability to record plane orientation allows access to information on symmetry of the unit cell, and in the context of laser compression, has been used successfully to infer strength\cite{Comley2013} and defect mediated lattice rotation\cite{Suggit2012}. However, as only plane orientation and not spacing can be determined from this technique, a full determination of volumetric compression, a key quantity for the interpretation of data, is lacking. 

Single Hit Energy-resolved Laue Diffraction (SHiELD) is an \textit{in-situ} white light Laue x-ray diffraction platform with the advantage that both the volume and aspect ratio of the unit cell can be measured from a single shot diffraction pattern. It utilises one or more CCD cameras operated in single-photon mode\cite{Stoeckl2004,King2005,Higginbotham2014} to record x-rays diffracted from a single crystal sample. As photons are detected individually by the CCD, their energy, and thus the spacing of the diffracting plane, can be recovered. By measuring photon energy for a small subset of Laue reflections one can fully supplement the cell aspect ratio information provided by the standard Laue technique to provide more complete strain state information for the sample.

The high signal to noise level associated with the Laue diffraction platform coupled with the sensitivity of single photon counting means that the technique is also well suited to high noise environments, a feature which may become important in the quest to obtain diffraction from materials at higher dynamic pressures, which in turn will necessitate greater laser ablation pressures, resulting in increased noise owing to x-rays emitted by the ablation plume.  However, the SHiELD technique affords recording of both the energy  and position on the detector of the observed photons.  For photons that have been diffracted from the sample there exists a correlation, owing to Bragg's law, between scattering angle and photon energy which does not exist for noise photons, thus allowing a means to separate signal from noise, even when both signal and noise photons are recorded at similar positions on the detector.   

\begin{figure} 
  \centering
  \includegraphics[width=0.48\textwidth]{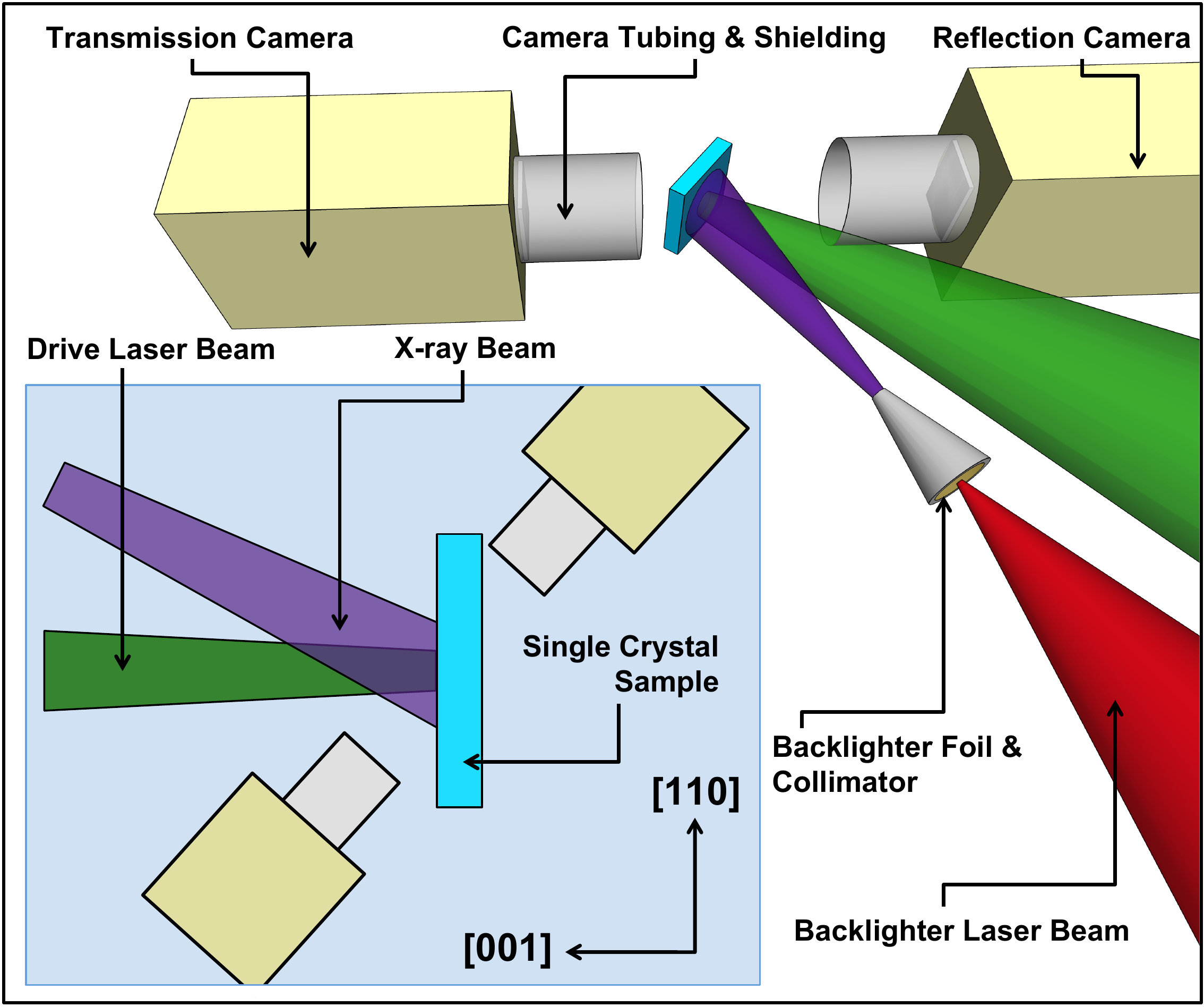}
  \caption{Schematic diagram of the experimental setup. Inset: Top-down view of the experiment highlighting orientation relative to crystal axes.}
  \label{fig:experimentdiagram}
\end{figure}

Here we present initial results from shock-compressed single-crystals of silicon that successfully demonstrate the technique.  Our observations using SHiELD are consistent with strong elastic waves propagating into the silicon sample, followed by a region that does not diffract efficiently in a specular manner.  As we discuss below, the results are consistent with previous observations of the response of silicon to shock compression on these nanosecond time-scales.

\section{Experiment}
The experiment was performed with the Janus laser at the Jupiter Laser Facility in Lawrence Livermore National Laboratory. A quasi-white light x-ray source was generated by irradiation of a mixed metal foil backlighter by a 2\,ns laser pulse at 527\,nm. The beam delivered up to 200\,J and was defocused to reach an intensity of $10^{14}\,\text{Wcm}^{-2}$, which has previously been shown to generate quasi-white light x-rays from 3\,-\,10 keV\cite{Suggit2010}. The x-rays were collimated using a molybdenum tube to limit the divergence of the x-ray beam, constraining the size of the x-ray spot to a diameter of $\approx$\,2.5\,mm on the samples, which are placed 4.2\,cm from the x-ray source. A schematic of the experimental setup is shown in figure \ref{fig:experimentdiagram}.

A second beam was used to shock compress single crystal samples, via laser-plasma ablation, using temporally square pulses 5\,-\,8\,ns long with up to 70\,J of energy. A random phase plate was used with the drive beam, creating a 1\,mm$^2$ square spot on the sample. This achieved intensities of up to 8$\times$10$^{11}$\,Wcm$^{-2}$ on target, producing a planar shock of up to 25\,GPa within the sample. During the experiment the x-ray backlighter beam was delayed relative to the drive beam, so that the x-rays could probe the sample at different stages of its compression.

Samples were 50\,${\mu}$m thick, $\left[001\right]$ oriented silicon single crystals with a 30\,${\mu}$m parylene-N ablator and a 0.2\,${\mu}$m aluminium flash layer. The aluminium layer is sufficiently thin such that diffraction from it was not expected and indeed not observed. The sample's rear surface velocity was measured with a velocity interferometer (VISAR)\cite{Celliers1998}, allowing for the determination of sample pressure and shock planarity. Two different synchronised VISAR measurements were made using different etalon thicknesses, achieving velocity per fringe values of 988 and 1731 ms$^{-1}$, in order to remove ambiguity due to large fringe shifts. 

Two Princeton Instruments MTE CCD cameras were used to detect x-rays diffracted from the sample. They were cooled to -30\,$\degree$C to minimise the number of dark counts generated in each pixel. Both cameras were shielded from optical light by thin Be windows and Al coated plastic tubing was used to restrict their field of view to only the target. 

For a diagonalised deformation tensor (i.e. one that is parameterised by a transverse and longitudinal strain with no off-diagonal elements), such as one would expect for a purely elastically compressed material, the cameras can be positioned such that they receive signal from Laue spots that do not move under compression. Specifically, these are spots whose associated reciprocal lattice vector lies either parallel or perpendicular to the compression axis. This greatly simplifies setup, allowing us to record diffraction from both ambient and compressed material simultaneously (assuming that the backlighter is timed such that the compression wave is only part way through the sample). For the [001] oriented silicon used here, we record diffraction from the (004) plane, which contains only information on strain along the compression axis, and (220), which is only sensitive to transverse strains. 

For the camera on the driven side of the sample (the reflection camera which records the (004) diffraction) x-rays from the target were attenuated by a 50\,${\mu}$m aluminium filter and 300\,$\mu$m PETE plastic placed at the end of the plastic tubing. This reduced the signal levels measured on the camera and significantly attenuated drive noise (which is typically of photon energy $<$\,4\,keV) from the sample in order to ensure the cameras operated in the single photon regime. The relatively large degree of filtering, compared with standard Laue diffraction, was primarily used to reduce the amount of diffraction signal. This is typical for a Laue diffraction pattern and indicates the technique's suitability for high noise environments. For the camera which recorded the (220) reflection  (the transmission camera positioned behind the sample) only a beryllium filter was used in order to maximise diffraction signal at the expense of increased sensitivity to drive noise.

\section{Results}

\begin{figure}  
  \centering
  \includegraphics[width=0.48\textwidth]{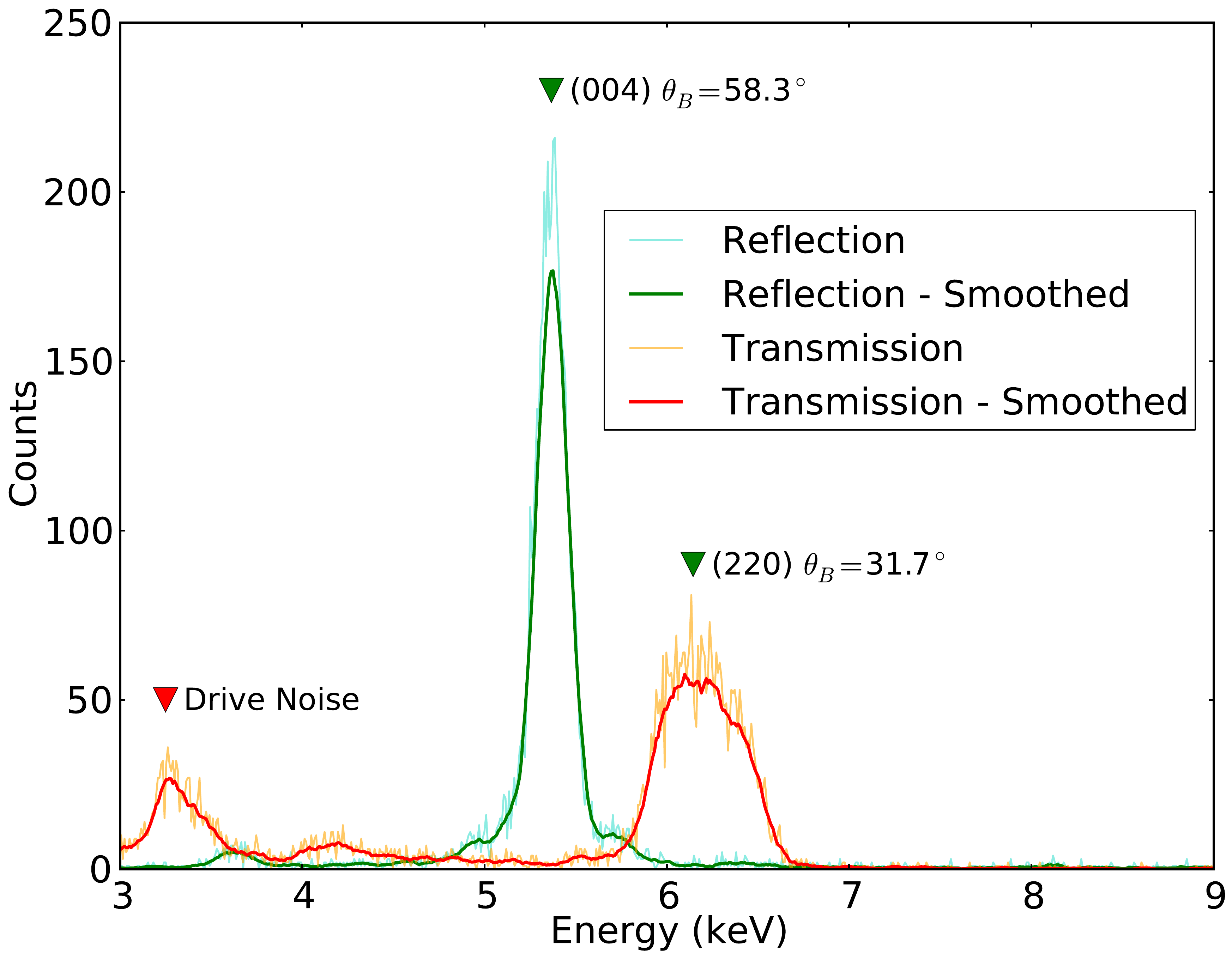} 
  \caption{Representative diffraction spectra measured by the two cameras from a single shot. The two Bragg angles indicated sum to 90$\degree$ which is consistent with diffraction from planes that are perpendicular to each other. The transmission camera measures a large amount of noise from ablation plasma due to having significantly less filtering. This signal is only present on shots with driven samples and the photons are randomly distributed across the entire the CCD which is consistent with noise processes.}
  \label{fig:SpectrumExample}
\end{figure}

The spectrum of photon energies measured from both cameras for a shot with a driven sample is shown in figure \ref{fig:SpectrumExample}. The data show a clear peak in the drive side camera spectrum for which the photon energy corresponds to diffraction from the (004) plane of silicon. A similar peak is seen in the transmission diffraction spectrum, where here the main peak corresponds to diffraction from the (220) plane.  Note that since the camera recording the (220) plane used less filtering, it is more sensitive to scattered background photons.  The peak of the photon energies are very close to those expected for an undriven silicon crystal sample.  This is because the size of the region on the crystal that can Laue-diffract x-rays (which is determined by the x-ray divergence) is larger than the region of the crystal within it which is shock-compressed (which is determined by the size of the drive beam). 

\begin{figure}
   \centering
   \includegraphics[width=0.48\textwidth]{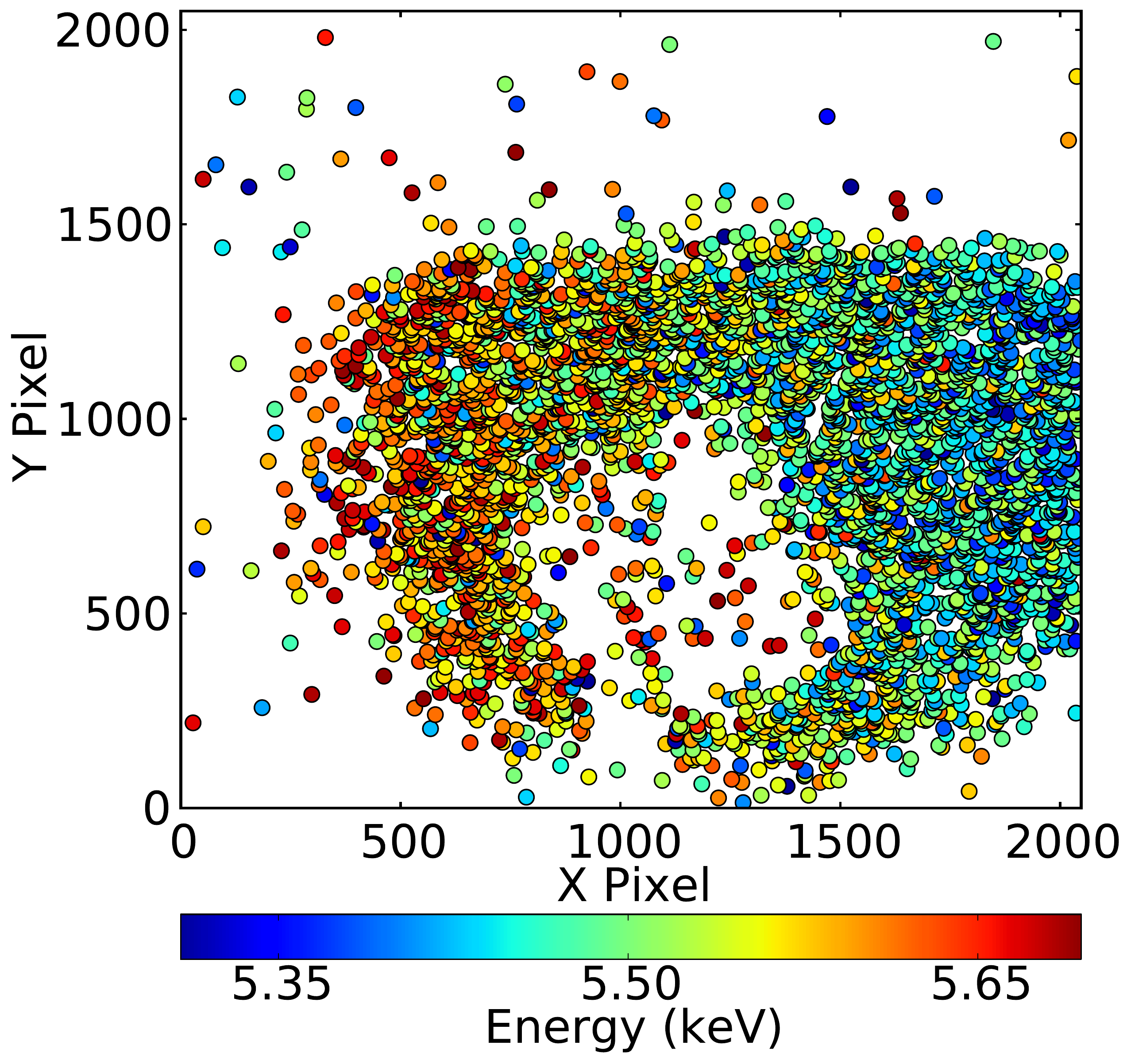} 
   \caption{(Color Online) Single photon events comprising the (004) diffraction peak plotted by pixel location. Each photon is coloured by its energy. Note the correlation between energy and spatial position within the diffraction spot which is absent for the noise photons lying outside it.}
   \label{fig:CCDExample}
\end{figure}

The position of the photons, colour-coded by energy, which comprise the (004) diffraction peak are plotted in figure \ref{fig:CCDExample} showing a variation in diffracted photon energy across the CCD due to the divergence of the x-ray source. The square drive spot is clearly evident in the centre of the diffraction pattern as the region where there is a reduced number of recorded photons.  We discuss below the reason for this reduction in the number of recorded photons within the drive spot, but for the moment note that this spatial resolution of the driven region not only allows us to verify the alignment between the x-rays incident on the crystal and the driven region, but furthermore means we can analyse the energy of only those photons that come from the drive region (or, of course, are noise, but may be discarded owing to poor correlation between position and energy). Additionally the relative backlighter fluence between shots can be quantitatively measured by measuring the number of diffracted photons per unit area in the undriven region of the diffraction pattern; a measurement which would otherwise require a separate spectrometer. The sharp cut-off of diffracted photons towards the top of the CCD is due to the edge of the single-crystal sample. Photons detected outside of these peaks are believed to be from noise sources, such as incoherent scatter of backlighter or drive plasma x-rays from the target chamber. This interpretation is supported by noting that for these photons we do not see any correlation between photon energy and spatial position on the CCD (i.e. scattering angle).

\begin{figure}
   \centering
   \includegraphics[width=0.48\textwidth]{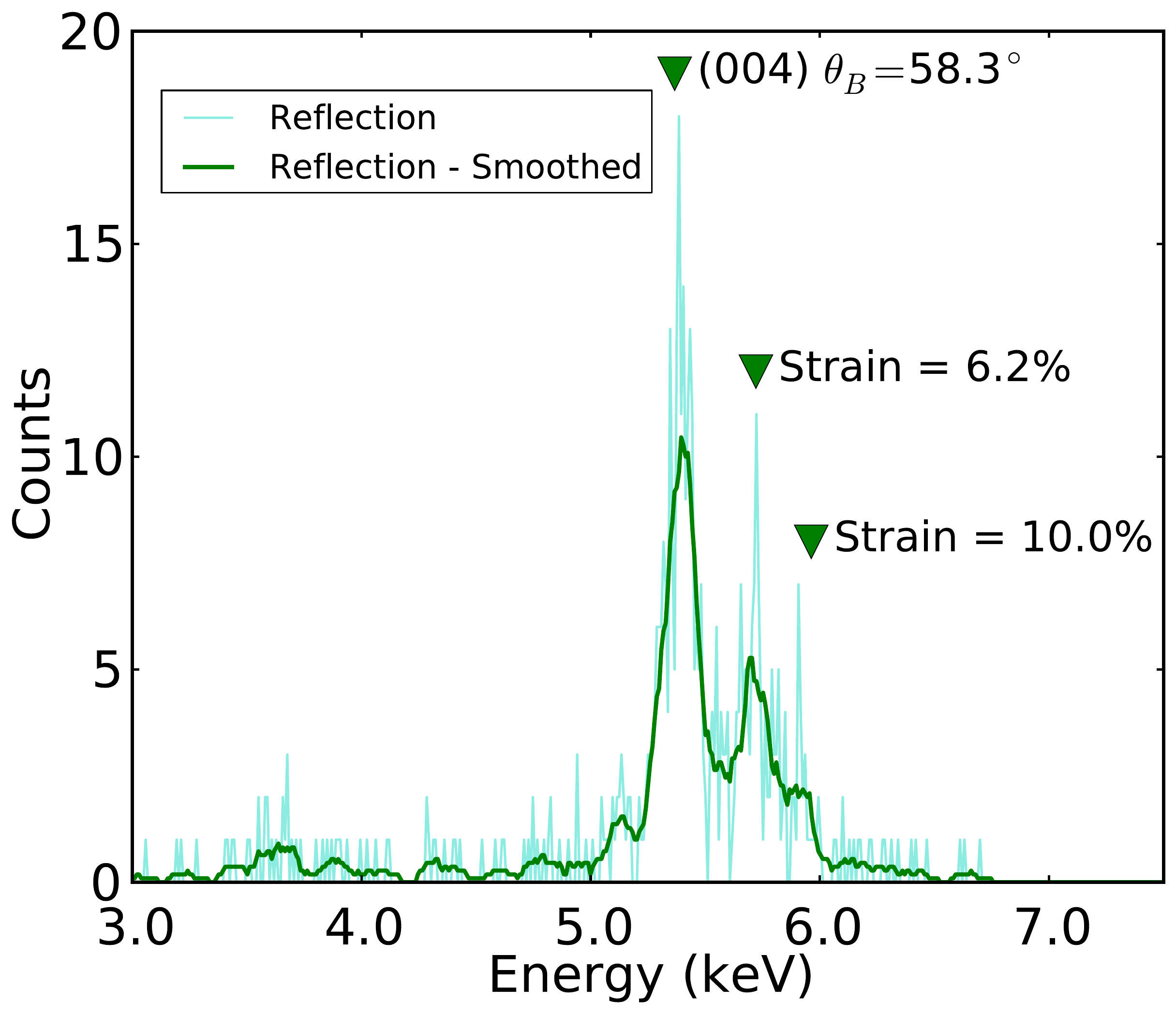} 
   \caption{(Color Online)Spectrum of photons taken from within the square driven region, corresponding to driven material, seen in figure \ref{fig:CCDExample}.}
   \label{fig:DrivenSpectrum}
\end{figure}

The spectrum of photon energies measured by the reflection camera, taken solely from the region of the CCD corresponding to diffraction from driven material, is shown in figure \ref{fig:DrivenSpectrum}. This spectrum shows the diffraction peak from uncompressed crystal, but also a second peak consistent with compressive strain of 6.2\,\% along the shock direction. There are also indications of elastic response from material at higher strains up to 11\,\% which is consistent with previous laser compression work where similar anomalous elastic strains are reported\cite{Loveridge-Smith2001}.
	
A similar analysis was performed on the energies of the diffracted photons recorded by the transmission camera (i.e. from the (220) planes). However no evidence of transverse strain was observed -  only uncompressed material. This observation is also consistent with previous in-situ diffraction work on silicon where no transverse strain was observed indicating a purely elastic response from the compressed material contributing to the observed diffraction \cite{Loveridge-Smith2001a,Loveridge-Smith2001}. This allows us to infer that the observed strains correspond to 6.2\% (and potentially 11\%) volumetric compression of the sample. Note that without the information about photon energy, which we glean from this single photon detection technique, the white light Laue technique would only enable us to determine  that a 6.2\% differential between longitudinal and transverse strains was present, not the absolute values of these strains.  Furthermore, even this information would require diffraction from planes with normals that are not parallel or perpendicular to the compression direction (e.g. in this case (111)).  
	
Note also that this technique retains a large degree of spatial resolution. Therefore, in the case of material undergoing small rotations, for which the diffraction will remain within the area of the CCD chip, energy, and thus strain can be resolved over multiple distinct structures, as has been seen in previous work employing Laue diffraction.\cite{Suggit2012}.

\begin{figure}   
   \centering
   \includegraphics[width=0.48\textwidth]{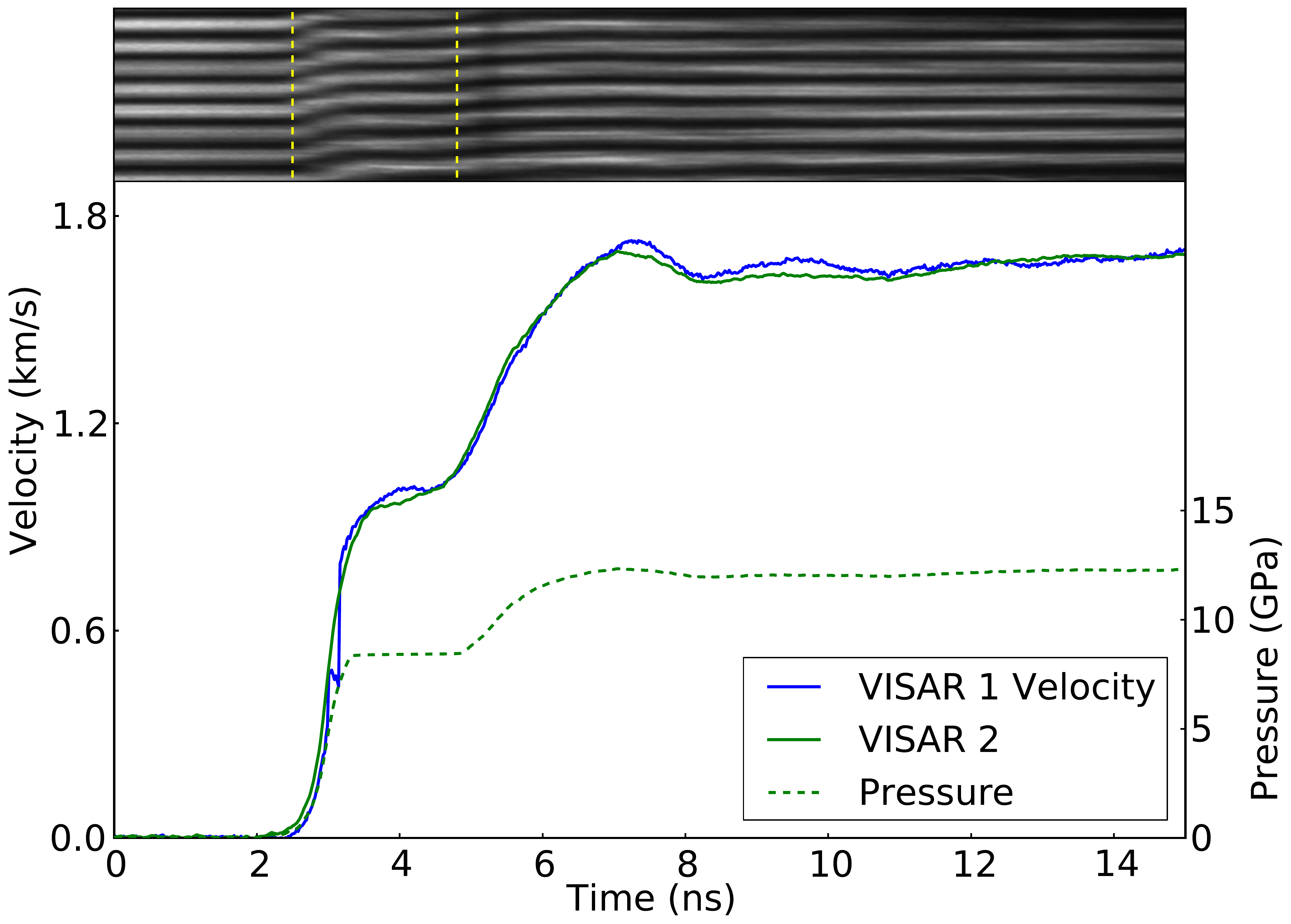} 
   \caption{Top: Streaked interferogram from VISAR 2. The vertical dashed lines indicate the arrival times of the two waves at the rear surface. Bottom: Typical VISAR velocity profile with clear two wave structure. There was similar agreement between both VISAR tracks for all shots. The inferred longitudinal stress has also been plotted.}
   \label{fig:VisarTrace}
\end{figure}

The velocity profiles which are shown in figure \ref{fig:VisarTrace} clearly indicate the presence of a two wave structure. The first can be identified as the elastic shock front reaching the rear surface, and corresponds to a free surface velocity of 1 kms$^{-1}$, which is consistent with a Hugoniot elastic limit of order 9.2\,GPa\cite{Gust1971}.
		
The cause of the second wave, corresponding here to a stress of 12.5 \,GPa, has previously been associated with plasticity or a phase change \cite{Smith2012,Mogni2014, Turneaure2007}, although there was no direct evidence of plasticity or a phase change in the diffraction data recorded this experiment.  We discuss this finding further below.

\section{Discussion}

It is clear that the intensity of the diffracted x-rays from  the driven region of the Laue spot shown in figure \ref{fig:CCDExample} is considerably reduced compared with that from the surrounding undriven material. This is not consistent with the response of the crystal to shock compression being purely elastic throughout, as elastic compression should not alter the orientation of the (004) planes.  Thus the diminution of the signal suggests that not all of the material is diffracting the incident x-rays in a specular manner.

Recent molecular dynamics simulations performed by Mogni et al. (2014) indicate that silicon may relax the high shear stresses experienced under shock compression by transforming to a mixed phase, with each phase being comprised of very small crystals, of order a few nm in dimension, and severely distorted and rotated, that will consequently not diffract efficiently in this geometry. \cite{Mogni2014}  This transformation wave follows the elastic wave, and thus diffraction from the elastically compressed material will be attenuated by this non-diffracting region, and if the rotation of the small crystallites in the transformation wave is sufficient, photons scattered from it would not be detected.  In this scenario, the attenuation of the signal in the drive spot shown in figure \ref{fig:CCDExample} would correspond to the non-diffracting region being of order 15-$\mu$m thick.  We find further support for this interpretation from the fact that the signal from the driven region was observed to get weaker at later times, but the number of photons recorded in the dataset collected was insufficient to determine an accurate velocity of the transformation wave, though in principle this should be possible in future experiments.

The camera locations used here, ideally suited to measuring elastic and plastic strains, may not be suited to larger scale lattice changes such as those due to change of phase or twinning. However using predictions of how a material may undergo such a change, for example from results of molecular dynamics simulations, judicious choice of camera location would enable measurements from materials exhibiting complex changes whilst under compression. Due to the small solid angle covered by a single CCD this could either require a large number of cameras, or opting to use CCDs with large chip dimensions.

\section{Conclusion}

Single Hit Energy-resolved Laue Diffraction (SHiELD) has been demonstrated as a technique for performing in-situ Laue x-ray diffraction with the additional advantage of being able to measure both unit cell volume and aspect ratio simultaneously. Two CCD cameras operated in single photon mode have been used to measure diffraction from laser compressed samples, simultaneously measuring transverse and longitudinal components of the deformation tensor. 

This technique has successfully measured diffraction signal from uniaxially compressed silicon indicating elastic compression and the experiment has shown that the additional wave following the elastic shock wave attenuates the diffraction signal from the elastically compressed material upstream but does not itself diffract efficiently in the Laue geometry. 

\section{Acknowledgements}

We kindly acknowledge the assistance of the Jupiter laser facility staff. AH and PS gratefully acknowledge support from AWE.  JSW and  MJS acknowledge support from EPSRC under grant number EP/J017256/1.  SP and PS acknowledge support from EPSRC.
\section{References}
\bibliographystyle{apsrev}
\bibliography{/Users/Shamim/Documents/library}

\end{document}